\documentclass[twocolumn]{aastex62}
\usepackage{natbib}
\usepackage{float}
\usepackage{amssymb}
\usepackage{amsmath}
\usepackage{comment}
\usepackage{rotating}
\usepackage{tabularx}

\begin{document}
\title{Optical color of Type Ib and Ic supernovae and implications for their progenitors}
\author{Harim Jin}
\affil{Department of Physics and Astronomy, Seoul National University, 08826, Seoul, South Korea}
\affil{Argelander Institute for Astronomy, University of Bonn, D-53121, Bonn, Germany}
\author{Sung-Chul Yoon}
\affiliation{Department of Physics and Astronomy, Seoul National University, 08826, Seoul, South Korea}
\affiliation{SNU Astronomy Research Center, Seoul National University, Seoul 08826, Korea}
\affiliation{Center for Theoretical Physics (CTP), Seoul National University, 08826, Seoul, South Korea}
\author{Sergei Blinnikov}
\affiliation{Kavli Institute for the Physics and Mathematics of the Universe (WPI), The University of Tokyo Institutes for Advanced Study, The University of Tokyo, 5-1-5 Kashiwanoha, Kashiwa, Chiba 277-8583, Japan}
\affiliation{NRC ``Kurchatov Institute"—ITEP, Kurchatov sq. 1, 123182  Moscow, Russia}
\affiliation{All-Russia Research Institute of Automatics (VNIIA), 127055 Moscow, Russia}

\begin{abstract}
Type Ib and Ic supernovae (SNe Ib/Ic) originate from hydrogen-deficient massive
star progenitors, of which the exact properties are still much debated.  Using
the SN data in the literature, we investigate the optical $B-V$ color of SNe
Ib/Ic at the $V-$band peak and show that SNe Ib are systematically bluer than
SNe Ic. We construct SN models from helium-rich and helium-poor progenitors of
various masses using the radiation hydrodynamics code STELLA and discuss
how the $B-V$ color at the $V-$band peak is affected by 
$^{56}$Ni to ejecta mass ratios, $^{56}$Ni mixing and
presence/absence of the helium envelope. 
We argue that the dichotomy in the amounts of helium in the progenitors 
plays the primary role in making the observed systematic color
difference at the optical peak, in favor of the most commonly invoked SN scenario that 
SNe Ib and SNe Ic progenitors are helium-rich and helium-poor, respectively.
\end{abstract}

\section{Introduction}\label{sect:introduction}
How different are Type Ib and Type Ic supernova (SN Ib/Ic) progenitors from
each other?  The absence of any noticeable hydrogen lines in SNe Ib/Ic spectra
implies that the hydrogen envelope of their progenitors is stripped off because
hydrogen cannot be easily hidden in the spectra even for a small
amount~\citep[e.g.,][]{Dessart2011}.  However, the absence of \ion{He}{1} lines
in SNe Ic spectra does not necessarily mean that their progenitors lack the
helium envelope. The formation of \ion{He}{1} lines requires non-thermal
processes \citep{Lucy1991} and a large amount of helium could be easily hidden
in the spectra if radioactive $^{56}$Ni were not present in the helium-rich
layer in the SN ejecta~\citep[e.g.,][]{Dessart2012}.  This makes it difficult
to observationally constrain the amount of helium retained in SN Ic
progenitors, obstructing our comprehensive understanding of different
evolutionary channels towards SNe Ib/Ic~\citep{Yoon2015, Yoon2017b}.  

Several authors have calculated SN Ib/Ic spectra from stripped-envelope progenitors
having different chemical compositions (i.e., different amounts of helium 
and various degrees of $^{56}$Ni mixing; 
\citealt{Dessart2012, Dessart2015, 
Dessart2020, Hachinger2012, Teffs2020, Williamson2020}).  These studies indicate
that while a large amount of helium ($M > \sim 1.0 M_\odot$) can be hidden
at around the optical peak without the presence of radioactive  $^{56}$Ni in
the helium-rich layer~\citep[see][]{Dessart2012}, only a small amount ($M \sim
0.1 M_\odot$, depending on the ejecta mass) can lead to formation of strong
\ion{He}{1} lines if $^{56}$Ni is sufficiently mixed into the helium-rich
layer.  

Recent analyses of a large set of SNe Ib/Ic spectra seem to imply
distinct progenitor chemical compositions. For example, a stronger and broader
\ion{O}{1} $\lambda7774$ absorption line and a broader \ion{Fe}{2} $\lambda5169$
line found in SNe Ic spectra than in SNe Ib,  disfavoring the existence of a
large amount of helium left in their progenitors \citep{Matheson2001, Liu2016,
Fremling2018}. The broader spectral lines might be also related to  higher
degrees of $^{56}$Ni mixing in SNe Ic than SNe Ib~\citep[cf.][]{Yoon2019},
which would make the early photospheric velocity faster \citep{Moriya2020}.
More recently, \citet{Shahbandeh2022} investigated a large set of near-infrared
spectra of stripped envelope SNe, finding that the \ion{He}{1}
$\lambda2.0581~\mu\mathrm{m}$ line of SNe Ic is systematically weaker than SNe
Ib. 

While distinct features are found in SNe Ib and Ic spectra, their light curves
are comparable in terms of the peak luminosity and the width. These observable
properties are related to the explosion parameters (i.e., ejecta mass
$M_\mathrm{ej}$, $^{56}$Ni mass $M_\mathrm{Ni}$, and kinetic energy
$E_\mathrm{k}$). Light curve analyses on large samples of SN Ib/Ic  do not lead
to a robust consensus on whether ordinary SNe Ib and Ic have systematically
different explosion parameters from each other, while broad-lined SNe Ic  have
higher kinetic energies and higher ejecta and $^{56}$Ni masses on average
\citep[e.g.,][]{Richardson2006, Drout2011, Cano2013, Taddia2015, Lyman2016,
Prentice2016, Taddia2018, Prentice2019, Barbarino2020, Zheng2022}. 

Optical color and its evolution reveal both similar and dissimilar properties
of SNe Ib/Ic. In the $V-R$ color of SNe Ib/Ic, a small scatter is observed at
10 days after the $V-$band maximum and is used to infer their host galaxy
reddening \citep{Drout2011, Stritzinger2018}. On the other hand, the early-time
color evolution of SNe Ib and Ic seems to show distinct features, implying
different degrees of $^{56}$Ni mixing in their ejecta \citep{Yoon2019}. 

In this study, we present another meaningful signature of different natures of
SNe Ib and Ic progenitors: optical color near the optical maximum. We show that
SNe Ib are systematically bluer than SNe Ic at the optical peak using SNe Ib/Ic data in the
literature and argue that the difference in the chemical structures of their
progenitors can explain such a color gap, using SN models calculated with the
radiation-hydrodynamics code STELLA. \citealt{Woosley2021}  also recently found
that the SN Ib/Ic color at 10 days after the $V$-band peak is systematically 
redder for helium-rich models than helium-poor models, in line with this study. 

The paper is organized as follows. We introduce our selected SN Ib/Ic sample
and present their $B-V$ color at the $V-$band peak in Section~\ref{sec:obs}.
Then we present our SN models newly constructed for this study in
Section~\ref{sec:mod}. In Section~\ref{sec:col}, we compare the models with the
observation and discuss the possible origins of the color difference. We
conclude the study in Section~\ref{sec:con}.

\section{Supernova sample} \label{sec:obs}

\begin{table*}[]
\begin{center}
\caption{List of our selected sample of SN Ib/Ic}\label{tab:obs}
\begin{tabular}{p{0.09\textwidth}p{0.035\textwidth}p{0.04\textwidth}p{0.04\textwidth}p{0.12\textwidth}p{0.10\textwidth}p{0.12\textwidth}p{0.06\textwidth}p{0.12\textwidth}p{0.04\textwidth}p{0.07\textwidth}p{0.03\textwidth}}
\hline \hline
Name & Type & $B_\mathrm{max}$ & $V_\mathrm{max}$ & Ref. & $E(B-V)_\mathrm{MW}$ & $E(B-V)_\mathrm{Host}$ & Method & Ref. & $M_\mathrm{Ni}$ & $M_\mathrm{ej}$ & Ref. \\ \hline
SN 1999ex & Ib & 17.44 & 16.60 & S02 & 0.02 & $0.28\pm0.04$ & SNIa & L16(S02) & 0.15 & 2.9 & L16 \\
SN 2004gq & Ib & 15.89 & 15.29 & D11, S18 & 0.0645 & 0.11$\pm$0.08 & Template& T18 & 0.11 & 3.4 & T18 \\
SN 2004gv & Ib & 17.68 & 17.25 & S18 & 0.0290 & 0.03$\pm$0.013 &Template& T18 & 0.16 & 3.4 & T18 \\
SN 2006ep & Ib & 18.31 & 17.40 & Bi14, S18 & 0.0319 & 0.12$\pm$0.01 & Template & T18 & 0.12 & 1.9 & T18 \\
SN 2006gi & Ib & 17.10 & 16.18 & E11 & 0.0248 & 0.098 & NaID  & E11 & 0.064 & 3.0 & E11 \\
SN 2006lc & Ib & 18.92 & 17.68 & Bi14, S18 & 0.0571 & 0.47$\pm$0.09& Template & T18 & 0.14 & 3.4 & T18 \\
SN 2007C & Ib & 17.25 & 15.98 & Br14, Bi14 & 0.0374 & 0.55$\pm$0.04 & Template& T18 & 0.07 & 6.2 & T18 \\
SN 2007kj & Ib & 18.14 & 17.64 & Bi14, S18 & 0.0713 & 0.00 & Template & T18 & 0.066 & 2.5 & T18 \\
SN 2007Y & Ib & 15.61 & 15.30 & Br14, S18 & 0.0190 & 0.00 & Template & T18 & 0.03 & 1.9 & T18 \\
SN 2008D & Ib & 18.51 & 17.33 & M08, Br14 & 0.0 & 0.60$\pm$0.20& NaID & L16(M09) & 0.09 & 2.9 & L16 \\
SN 2009jf & Ib & 15.58 & 15.08 & S11 & 0.112 & 0.005 $\pm$ 0.05& NaID & L16(V11) & 0.24 & 4.7 & L16 \\
SN 2012au & Ib & 14.02 & 13.51 & M13 & 0.043 & 0.02 $\pm$ 0.01 & NaID & M13 & 0.3 & $^{*}$4(3-5) & M13 \\
SN 2014C & Ib & 16.04 & 14.93 & Br14 & 0.08 & 0.67 $\pm$ 0.08 & NaID & M17(M15) & 0.15 & 1.7 & M17 \\
SN 2015ah & Ib & 17.10 & 16.50 & P19 & 0.071 & 0.02 $\pm$ 0.01 &NaID & P19 & 0.092 & 2.0 & P19 \\
SN 2015ap & Ib & 15.71 & 15.20 & P19 & 0.037 & 0.00 $\pm$ 0.00  & NaID & P19 & 0.12 & 1.8 & P19 \\
iPTF13bvn & Ib & 15.91 & 15.21 & F16 & 0.0278 & 0.0437 & NaID & L16(F14) & 0.06 & 1.7 & L16 \\ \hline
SN 1994I & Ic & 13.83 & 12.87 & T93 & 0.0308 & 0.269 $\pm$ 0.16 & NaID  & L16(R96), G17 & 0.07 & 0.6 & L16 \\
SN 2004aw & Ic & 18.11 & 17.12 & Bi14 & 0.021 & 0.35 $\pm$ 0.10 & NaID&  L16(T06) & 0.20 & 3.3 & L16 \\
SN 2004dn & Ic & 18.68 & 17.32 & D11, G05 & 0.048 & 0.52 $\pm$ 0.13 & $V-R$ & L16(D11) & 0.16 & 2.8 & L16 \\
SN 2004fe & Ic & 17.55 & 16.88 & D11, Bi14, S18 & 0.0216 & 0.00 & Template & T18 & 0.1 & 2.5 & T18 \\
SN 2004gt & Ic & 16.36 & 15.40 & S18 & 0.0410 & 0.43$\pm$0.06 & Template  & T18 & 0.16 & 3.4 & T18 \\
SN 2005aw & Ic & 17.24 & 15.99 & S18 & 0.0542 & 0.46$\pm$0.04 & Template & T18 & 0.17 & 4.3 & T18 \\
SN 2007gr & Ic & 13.48 & 12.88 & H09 & 0.062 & 0.03$\pm$0.018 & NaID & L16(H09) & 0.08 & 1.8 & L16 \\
SN 2007hn & Ic & 19.17 & 18.28 & S18 & 0.0710 & 0.134$\pm$0.03 & Template  & T18 & 0.25 & 1.5 & T18 \\
SN 2011bm & Ic & 17.11 & 16.52 & V12 & 0.032 & 0.032 & NaID & L16(V12) & 0.62 & 10.1 & L16 \\
SN 2013F & Ic & 19.15 & 17.13 & P19 & 0.018 & 1.4 $\pm$ 0.2 & NaID & P19 & 0.15 & 1.4 & P19 \\
SN 2013ge & Ic & 15.54 & 14.76 & D16 & 0.020 & 0.047  &NaID & D16 & 0.12 & ${^*}$2.5(2-3) & D16 \\
SN 2014L & Ic & 16.22 & 15.04 & Z18 & 0.04 & 0.63 $\pm$ 0.11 & NaID & Z18 & 0.075 & 1.0 & Z18 \\
SN 2016iae & Ic & 16.14 & 14.99 & P19 & 0.014 & 0.65 $\pm$ 0.20 &NaID & P19 & 0.13 & 2.2 & P19 \\
SN 2016P & Ic & 17.41 & 16.62 & P19 & 0.024 & 0.05 $\pm$ 0.02  & NaID& P19 & 0.09 & 1.5 & P19 \\
SN 2017ein & Ic & 16.01 & 15.26 & V18 & 0.019 & 0.40 $\pm$ 0.06 & NaID & X19 & 0.13 & 0.9 & X19 \\
SN 2020oi & Ic & 14.57 & 13.82 & R20 & 0.0227 & 0.00 & NaID & R21 & 0.07 & 0.7 & R21 \\
LSQ14efd & Ic & 19.79 & 18.96 & B17 & 0.0376 & 0.00 $\pm$ 0.0015 & NaID& B17 & 0.25 & 2.5 & J21 \\ \hline
\end{tabular}
\end{center}
\tablecomments{The first five columns give the name, SN subtype, $B-$band magnitude obtained at the $V$-band peak, $V-$band peak magnitude, and the references from which the photometric data are collected. The sixth, seventh, eighth, nineth columns are $E(B-V)_\mathrm{MW}$, $E(B-V)_\mathrm{Host}$, the adopted method for inferring host extinction and their references. The last three columns give the inferred values of $M_\mathrm{Ni}$, $M_\mathrm{ej}$, and their references. Ejecta masses with asterisks are middle values chosen from given ranges. References are abbreviated as follows. T93: \citet{Tsvetkov1993}, R96: \citet{Richmond1996}, S02: \citet{Stritzinger2002}, G05: \citet{GalYam2005}, T06: \citet{Taubenberger2006}, M07: \citet{Modjaz2007}, M08: \citet{Mazzali2008}, H09: \citet{Hunter2009}, M09: \citet{Modjaz2009}, D11: \citet{Drout2011}, E11: \citet{Elmhamdi2011}, S11: \citet{Sahu2011}, V11: \citet{Valenti2011}, V12: \citet{Valenti2012}, M13: \citet{Milisavljevic2013}, Bi14: \citet{Bianco2014}, Br14: \citet{Brown2014}, F14: \citet{Fremling2014}, M15: \citet{Milisavljevic2015}, D16: \citet{Drout2016}, F16: \citet{Folatelli2016}, L16: \citet{Lyman2016}, B17: \citet{Barbarino2017}, G17: \citet{Guillochon2017}, S18: \citet{Stritzinger2018},  T18: \citet{Taddia2018}, V18: \citet{VanDyk2018}, Z18: \citet{Zhang2018}, P19: \citet{Prentice2019}, X19: \citet{Xiang2019}, R21: \citet{Rho2021}, J21: \citet{Jin2021}}
\end{table*}

\begin{figure}
\plotone{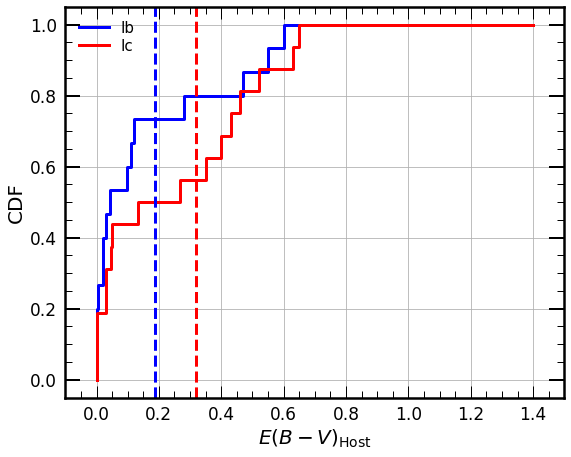}
\plotone{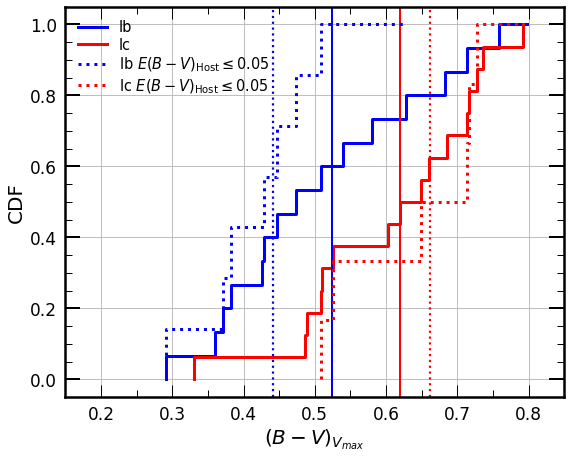}
\plotone{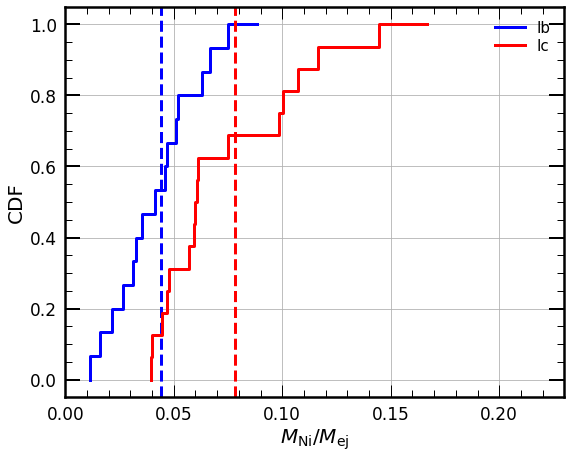}
\caption{\emph{Top panel:} Cumulative distribution of $E(B-V)_\mathrm{Host}$ (top), the $B-V$ color at the $V$-band peak after 
the host extinction correction (middle),  $M_\mathrm{Ni} / M_\mathrm{ej}$ (bottom) of our SN Ib/Ic sample.
The dotted line in the middle panel is the cumulative distribution of $(B-V)_{V_\mathrm{max}}$ for the case of minimal reddening  (i.e.,  $E(B-V)_\mathrm{Host} \le 0.05$).
The vertical lines indicate the mean value of each distribution. 
See the text for more details.}
\label{fig:obs}
\end{figure}

The majority of our sample SN Ib/Ic photometric data are collected from the
Open Supernova Catalog (OSC) \citep{Guillochon2017}. SNe Ib/Ic that have
more than 30 photometric data points are selected. We exclude superluminous SNe
Ic, broad-lined SNe Ic, and Ca-rich SNe Ib to focus our discussion on 
ordinary SNe Ib/Ic. Then SNe Ib/Ic that have the main peak information in the
$V$-band and have more than three $B$-band and $V$-band data points within $\pm
5$ days with respect to the $V$-band peak epoch are selected to obtain a
reasonably good $B-V$ color estimate at the $V$-band peak.

The host galaxy extinction for SNe Ib/Ic is usually non-negligible due to their
dusty environments. 
In Table~\ref{tab:obs}, we present the host extinction value of
$E(B-V)_\mathrm{Host}$ extracted from the literature and the corresponding
method to determine it for each SN in the 7th and 8th columns respectively,
along with the references in the 9th column where details on the host
extinction estimates can be found. 
The host extinction for SN 1999ex was obtained from the light curves 
of the Type Ia SN 1999ee, which occurred in the same host galaxy, assuming that it is not
much different from that of SN 1999ee~\citep{Stritzinger2002}. For SN 2004dn,
\citet{Drout2011} use the small scatter in the $V-R$ color of SN Ib/Ic at 10
days after the $V$-band peak and the Galactic extinction law to obtain the host
extinction. 
For the rest of the sample, either the `NaID' or `Template' method was used
as indicated in the table. 
Here, the NaID method denotes the conventional way
of using the equivalent width of Na I D absorption
lines~\citep[e.g.,][]{Munari1997, Poznanski2012}.  The `Template' method means the new method
introduced by \citet{Stritzinger2018}. These authors constructed a post-maximum
color template for each subtype of SESNe (i.e., SN IIb, Ib and Ic) 
from the optical and near-infrared light curves of 3
minimally reddened SNe of each subtype  and use them to infer the host extinctions of other
SNe. 

In the top panel of Figure~\ref{fig:obs}, cumulative
distributions of $E(B-V)_\mathrm{Host}$ are presented for both SNe Ib and Ic.
The overall host extinction is larger for SNe Ic
($\overline{E(B-V)}_\mathrm{Host}$= 0.32) than for SNe Ib
($\overline{E(B-V)}_\mathrm{Host}$= 0.19). This implies different progenitor
environments for SNe Ib and Ic: SNe Ic seem to originate from  more dusty
environments than SNe Ib, where more active star formation is expected.

Corrected for both the foreground extinction and the host galaxy extinction,
the $B-V$ color at the $V-$band peak, $(B-V)_\mathrm{Vmax}$, is obtained and
the corresponding cumulative distribution is presented in the middle panel of
Figure~\ref{fig:obs} with the solid lines.  It is observed that SNe Ib are
systematically bluer ($\overline{(B-V)}_\mathrm{Vmax}$= 0.52) than SNe Ic
($\overline{(B-V)}_\mathrm{Vmax}$= 0.62), with an average color difference of
$\Delta \overline{(B-V)}_\mathrm{Vmax}$ = 0.10 despite the fact that
systematically larger extinction corrections are applied to SNe Ic than SNe Ib.
The caveat is that there exist great uncertainties in the host extinction
estimate.  \citet{Munari1997} show that the NaID method is less reliable for
$E(B-V) \gtrsim 0.15$ because the Na I D lines are saturated. 
There are also multiple sources of uncertainties in this method,
including possible diverse dust properties of SN host galaxies and the effects
of the spectral resolution~\citep[e.g.,][]{Poznanski2011}.  The post-maximum
color templates by \citet{Stritzinger2018} are based on a sample of small size
(3 for each subtype) and the applicability of these templates needs to be
further investigated with a larger sample of minimally reddened SESNe.
For this reason we also present the cumulative distributions of
$\overline{(B-V)}_\mathrm{Vmax}$ only for SNe Ib/Ic having a very small
extinction(i.e., $E(B-V)_\mathrm{host} \le 0.05$) with the dotted lines (8 SNe Ib and 7 SNe Ic).  
With this minimally reddened sample,
the systematic color difference between SNe Ib and Ic becomes even more significant
(i.e., $\Delta \overline{(B-V)}_\mathrm{Vmax}$= 0.22). We conclude that this 
color difference probably reflects intrinsic properties of SNe Ib/Ic.

The inferred values of $M_\mathrm{Ni}$ and $M_\mathrm{ej}$ of the same SNe are
also collected from the literature as indicated in Table~\ref{tab:obs}, and
their ratios are presented in the bottom panel of Figure~\ref{fig:obs}.  SNe Ic
have a systematically higher $^{56}$Ni mass to ejecta mass ratio
($\overline{M_\mathrm{Ni} / M_\mathrm{ej}}$= 0.08) than SNe Ib
($\overline{M_\mathrm{Ni} / M_\mathrm{ej}}$= 0.04).  
As discussed below, a
higher $M_\mathrm{Ni}/M_\mathrm{ej}$ would lead to a bluer color for a given SN
ejecta property, meaning that the systematic redder color of SNe Ic than SNe Ib
could not be attributed to different $M_\mathrm{Ni}/M_\mathrm{ej}$ ratios.  The
caveat is that we find great uncertainties in the estimates of $M_\mathrm{Ni}$
and $M_\mathrm{ej}$  in the literature.  For example, in the case of SN 2007C,
we get $M_\mathrm{Ni} = 0.18~M_\odot$ \& $M_\mathrm{ej} = 1.83~M_\odot$ in
\citet{Cano2013} and $M_\mathrm{Ni} = 0.07 M_\odot$ \& $M_\mathrm{ej} =
6.2~M_\odot$  in \citet{Taddia2018}.  In the discussion below, therefore, we do
not try to reproduce observed light curves of individual SNe but focus on
general features of our models and qualitative comparison with the
observation.

\section{Supernova models} \label{sec:mod}

\subsection{Methods and physical assumptions}

We construct SN models from helium-rich and helium-poor progenitors for various
$M_\mathrm{ej}$, different amounts of $^{56}$Ni, and different $^{56}$Ni
distributions in the SN ejecta to explore the effects of  these factors on the
$B-V$ color.

\begin{figure}
    \centering
    \includegraphics[width=0.45\textwidth]{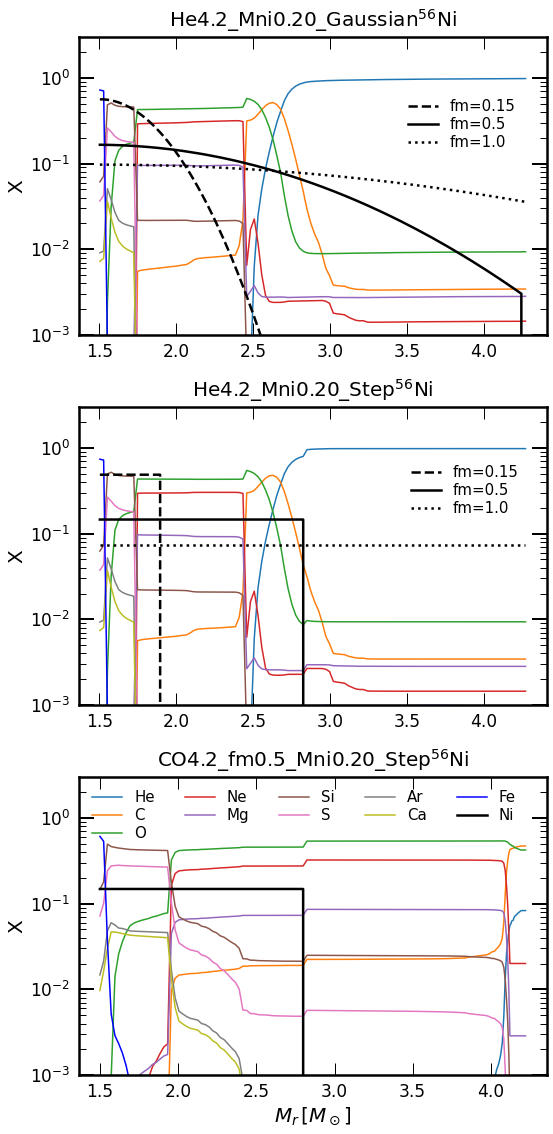}
\caption{Mass fractions of different chemical elements in the He4.2 model with Gaussian $^{56}$Ni distributions (top), with step $^{56}$Ni distributions (middle), and in the CO4.2 model with a step $^{56}$Ni distribution (bottom).
\label{fig:abn}} 
\end{figure}

\begin{table*}[]
\caption{Progenitor model properties}\label{tab:mod}
\begin{center}
\begin{tabular}{cccccccc|cccccccc}
\hline \hline
Name & $M_\mathrm{ej}$ & $R$ & $M_\mathrm{CO}$ & $Y_\mathrm{s}$ & $m_\mathrm{He}$ & $M_\mathrm{Fe}$ & $M_\mathrm{ext}$ & Name & $M_\mathrm{ej}$ & $R$ & $M_\mathrm{CO}$ & $Y_\mathrm{s}$ & $m_\mathrm{He}$ & $M_\mathrm{Fe}$ & $M_\mathrm{ext}$ \\
 & {[}$M_\odot${]} & {[}$R_\odot${]} & {[}$M_\odot${]} &  & {[}$M_\odot${]} & {[}$M_\odot${]} & {[}$M_\odot${]} &  & {[}$M_\odot${]} & {[}$R_\odot${]} & {[}$M_\odot${]} &  & {[}$M_\odot${]} & {[}$M_\odot${]} & {[}$M_\odot${]} \\ \hline
He3.1 & 1.73 & 31.66 & 1.56 & 0.98 & 1.43 & 1.30 & 0.018 & CO3.2 & 1.77 & 0.20 & 3.18 & 0.12 & 0.04 & 1.39 & 0.018 \\
He3.5 & 2.06 & 4.77 & 2.02 & 0.98 & 1.44 & 1.41 & 0.022 & CO3.6 & 2.05 & 0.21 & 3.61 & 0.12 & 0.07 & 1.53 & 0.021 \\
He3.9 & 2.40 & 6.73 & 2.17 & 0.98 & 1.66 & 1.44 & 0.024 & CO3.9 & 2.49 & 0.77 & 3.92 & 0.49 & 0.10 & 1.41 & 0.024 \\
He4.2 & 2.76 & 2.45 & 2.63 & 0.98 & 1.58 & 1.45 & 0.027 & CO4.2 & 2.72 & 0.22 & 4.19 & 0.08 & 0.06 & 1.45 & 0.027 \\
He5.3 & 3.75 & 0.87 & 3.95 & 0.82 & 0.63 & 1.50 & 0.038 & CO5.3 & 3.76 & 0.59 & 5.13 & 0.30 & 0.22 & 1.46 & 0.038 \\
He5.6 & 4.10 & 1.62 & 3.64 & 0.98 & 1.62 & 1.48 & 0.041 & CO5.7 & 4.08 & 0.20 & 5.56 & 0.30 & 0.17 & 1.63 & 0.041 \\ \hline
\end{tabular}
\end{center}
\tablecomments{$M_\mathrm{ej}$: ejecta mass; $R$: progenitor radius; $M_\mathrm{CO}$: mass of the helium-deficient core of which the helium mass fraction is lower than 0.2; $Y_\mathrm{s}$: surface helium mass fraction; $m_\mathrm{He}$: total helium mass; $M_\mathrm{Fe}$: iron core mass; $M_\textrm{ext}$: mass of the external material attached to the progenitor.}
\end{table*}



We still do not fully understand evolutionary paths of massive stars
towards SNe Ib or Ic and their mass-loss history \citep[see][for a detailed
discussion]{Yoon2017b}.  In this study, we do not aim to investigate pre-SN
evolution but instead consider diverse physical properties of SN Ib/Ic
progenitors at the pre-SN stage to investigate their impact on the supernova
light curves and colors.  For this purpose, we adopt different mass-loss rates
from helium stars to construct both helium-rich and helium-poor models having
various final masses as explained below. All the progenitor models have the
solar metallicity (i.e., $Z=0.02$) and are evolved until the infall velocity of
the iron core exceeds 1000~$\mathrm{km~s^{-1}}$ using the MESA
code~\citep{Paxton2011, Paxton2013, Paxton2015, Paxton2018, Paxton2019}.  
 
The helium-rich progenitor models  contain more than 0.6$M_\odot$ of helium (He
models) and the helium-poor progenitor models contain less than 0.2$M_\odot$ of
helium (CO models). Each progenitor model is named to indicate their
progenitor type and total mass. E.g., He3.1 refers to a helium-rich progenitor
with its total mass of 3.1~$M_\odot$. We use the same notation when referring
to the SN model from the respective progenitor model.  See Table~\ref{tab:mod}
for the details on the progenitor properties.

He3.1 and He3.9 are taken from the binary models Sm11p200d and Sm15p50 of
\citet{Yoon2017a}, respectively. He3.5, He4.2 and He5.6 are obtained by
evolving pure helium star models having initial masses of 4.0, 5.0, and 7.0
$M_\odot$ using the Wolf-Rayet mass-loss rate prescription by
\citet{Nugis2000}. He5.3 is obtained by evolving a 9.0~$M_\odot$ helium star
with the Wolf-Rayet mass-loss rate prescription by \citet{Yoon2017b}.  CO3.2,
CO3.6 and CO4.2 are obtained by evolving 7.0, 8.0, and 10.0 $M_\odot$ helium
stars with the Yoon mass-loss prescription until core helium exhaustion, and
with an artificially enhanced mass-loss rate (i.e., 500 times the standard Nugis \& Lamers rate)
thereafter until core-collapse.  CO3.9 and CO5.9
models are constructed in a similar fashion by evolving 7.0 and 10.0 $M_\odot$
helium stars but the Nugis \& Lamers mass-loss prescription is used during the
core-helium burning phase.  CO5.3 is calculated with a 11.0~$M_\odot$ helium
star model using the Yoon mass-loss prescription throughout the whole
evolution. Some example MESA inlist files for constructing these progenitor models can be found at
ZENODO: \dataset[10.5281/zenodo.7797106]{https://doi.org/10.5281/zenodo.7797106}.

The final masses span $M_\mathrm{tot}=3.1\cdots5.7M_\odot$, and the
corresponding ejecta masses span $M_\mathrm{ej}=1.7\cdots4.1M_\odot$ with the
assumption of this study that the mass cut in SN explosion is located at the
outer boundary of the iron core.  This range encompasses a large fraction of
the inferred ejecta masses of $M_\mathrm{ej} \approx 1.0 \cdots 5.0 M_\odot$ of
SNe Ib/Ic \citep[e.g.,][]{Drout2011, Cano2013, Lyman2016, Taddia2018,
Zheng2022}. An external material having a mass of about $1\%$ of the ejecta
mass and an extent of $10^{14}$cm is attached to each progenitor model to avoid
acceleration of the forward shock to a relativistic velocity because the
relativistic effects cannot be properly handled in the version of the STELLA
code that is used for this study \citep[see][for more disucssion on
this.]{Yoon2019}. This external matter does not affect the light curve
after about one day from the shock breakout, thus not affecting our focus of the study.

We use the STELLA code that is a one-dimensional
multi-group radiative-hydrodynamics code for calculating SN light curves in
multi-bands. The SN explosion is simulated by a thermal bomb at the mass cut, and
time-dependent radiative transfer equations and hydrodynamics equations are
solved simultaneously for approximately 100 frequency bins. For more detailed
information, see \citet{Blinnikov2011}. See also \citet{Yoon2019} for a recent
example of SNe Ib/Ic models calculated with the STELLA code.

We consider two kinetic energies, four $^{56}$Ni masses, and six $^{56}$Ni
distributions to construct SN models for a given progenitor model. Kinetic
energies of 1B and 2B are obtained by adjusting the explosion energy.
We do not calculate explosive nucleosynthesis but instead
radioactive $^{56}$Ni is artificially introduced into the ejecta as in
\citet{Yoon2019}.   The considered $^{56}$Ni masses are
$M_\mathrm{Ni}$=0.07$M_\odot$, 0.14$M_\odot$, 0.20$M_\odot$, and 0.25$M_\odot$.
These sets of kinetic energies and $^{56}$Ni masses cover most of the observed
values of ordinary SNe Ib/Ic \citep[e.g.,][]{Drout2011, Cano2013, Lyman2016,
Taddia2018, Zheng2022}. 

The $^{56}$Ni mass fraction in the SN ejecta is set to follow either a step
distribution or a Gaussian distribution. For a step distribution, we adopt
$X_\textrm{Ni}(M_r)=\frac{M_\textrm{Ni}}{f_\textrm{m}(M_\textrm{tot}-M_\textrm{cut})}$
for
$M_\textrm{cut}<M_r<f_\textrm{m}(M_\textrm{tot}-M_\textrm{cut})+M_\textrm{cut}$
and $X_\textrm{Ni}(M_r)=0$ elsewhere. Here, $M_r$ is the mass coordinate,
$M_\textrm{cut}$ the mass cut which corresponds to the outer boundary of the
iron core, $M_\textrm{tot}$ the total mass of the progenitor, and
$f_\textrm{m}$ is a free parameter that controls the shape of the $^{56}$Ni distribution
in the ejecta. For a Gaussian distribution, we adopt
$X_\textrm{Ni}(M_r)=A\textrm{exp}\left(-\left[\frac{M_r-M_\textrm{cut}}{f_\textrm{m}(M_\textrm{tot}-M_\textrm{cut})}\right]^2\right)$
where $A$ is the normalization factor. In the study, $f_\textrm{m}$=0.15, 0.5,
and 1.0 are considered for both step and Gaussian $^{56}$Ni distributions to
account for the uncertainty in  $^{56}$Ni mixing.  The value of
$f_\textrm{m}$=0.15 ($f_\textrm{m}$=0.5) corresponds to a weak (moderate)
$^{56}$Ni mixing. The value of  $f_\textrm{m}$=1.0 means complete or almost
full $^{56}$Ni mixing for the step and Gaussian distributions, respectively.
In Figure~\ref{fig:abn}, we present chemical profiles in the ejecta of He4.2
and CO4.2 models for various $^{56}$Ni distributions, as an example.  Note that
the  gamma-ray transfer in STELLA is treated in one-group approximation
calibrated against full Monte-Carlo transport and, in principle,  the deposition of the gamma-ray energy 
may take place far away from the birthplace of gamma photons in $^{56}$Ni and $^{56}$Co decays.  
In practice this non-local heating occurs only at late phases when the mean free path 
of gamma photons becomes large due to low density. In this study, however, 
we focus on relatively early phases where the mean free path of gamma photons
is short and the heating is effectively local. 

\subsection{Model color distributions} \label{subsec:mod}

\begin{figure*}
\epsscale{0.85}
\plotone{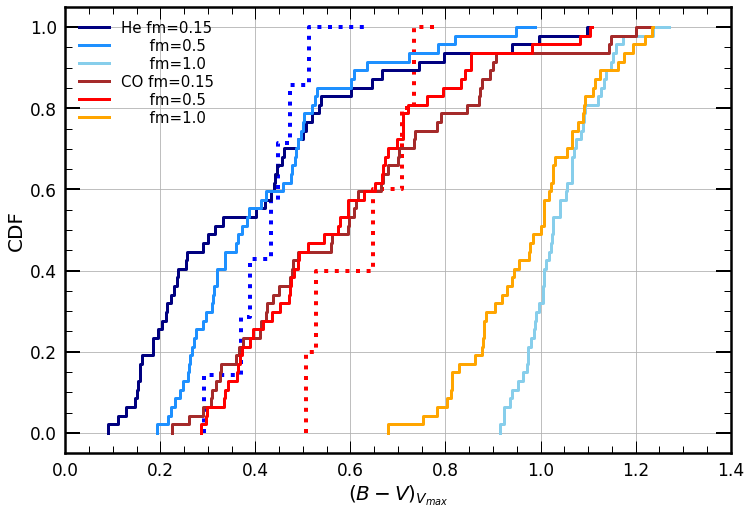}
\plotone{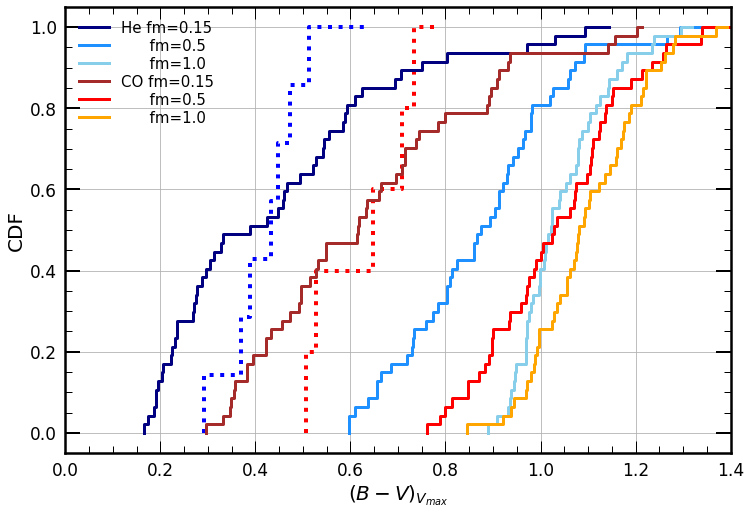}
\caption{\emph{Upper panel:} Cumulative distributions of $(B-V)_\mathrm{Vmax}$ of the STELLA models with the step $^{56}$Ni distributions.  
The observed distributions of our minimally reddened (i.e., $E(B-V)_\mathrm{host} \le 0.05$) SN Ib and Ic sample are given by the dotted blue and red lines. The average value of each case is given in Table~\ref{tab:avg}. \emph{Lower Panel:} Same as in the upper panel, but for the Gaussian $^{56}$Ni distributions.
}
\label{fig:bvdist}
\end{figure*}


\begin{table}[]
\begin{center}
\caption{Average $(B-V)_\mathrm{Vmax}$ values}\label{tab:avg}
\begin{tabular}{c|cc|cc}
\hline \hline 
$f_\mathrm{m}$ & \multicolumn{2}{c|}{Step} &  \multicolumn{2}{c}{Gauss}  \\
 & He & CO &  He & CO \\ \hline
0.15 & 0.40 & 0.61 &  0.45 & 0.65   \\
0.5 & 0.43 & 0.59 &  0.88 & 1.04  \\
1.0 & 1.04 & 0.99 &  1.05 & 1.10 \\ \hline
\end{tabular}
\end{center}
\tablecomments{Step: the average  $(B-V)_\mathrm{Vmax}$  values of He and CO models with the step $^{56}$Ni distribution,  Gauss: the average  $(B-V)_\mathrm{Vmax}$  values of the He and CO models with the Gaussian $^{56}$Ni distribution}
\end{table}

The cumulative distributions of $(B-V)_\mathrm{Vmax}$ predicted by the models
are presented in Figures~\ref{fig:bvdist} for both step and Gaussian $^{56}$Ni distributions. 
Although the SN parameters (i.e., $E_\mathrm{K}$, $M_\mathrm{ej}$ and
$M_\mathrm{Ni}$) of our models are chosen to reproduce ordinary SNe Ib/Ic, 
a precise quantitative comparison of the predicted $(B-V)_\mathrm{Vmax}$
with the observation would require a proper consideration of  the
exact distributions of these parameters within the selected SN Ib/Ic observation
sample.  However, there exist great uncertainties in the observationally
inferred values of these parameters as discussed above and we limit our
discussion to a qualitative comparison between the model prediction and the
observed values of $(B-V)_\mathrm{Vmax}$.

The He models with the step $^{56}$Ni distributions have a bluer color
($\overline{(B-V)}_\mathrm{Vmax} = 0.40 - 0.43$) than the CO models
($\overline{(B-V)}_\mathrm{Vmax} \sim 0.60$), except for the fully mixed case
($f_\mathrm{m}$=1.0) that leads to $\overline{(B-V)}_\mathrm{Vmax} \simeq 1.0$
for both He and CO models.  The systematic color difference between He and CO
models is $\Delta \overline{(B-V)}_\mathrm{Vmax} \approx$ 0.21 (0.16) for
$f_\mathrm{m}$=0.15 ($f_\mathrm{m}$=0.5), which is comparable to the observed
color difference between SNe Ib and Ic (i.e., $\Delta
\overline{(B-V)}_\mathrm{Vmax} =$ 0.10 and 0.22 for the full and minimally reddened
samples, respectively).

The He models with the Gaussian $^{56}$Ni distributions also have a bluer color
($\overline{(B-V)}_\mathrm{Vmax} \approx$ 0.45 \& 0.88) than CO models
($\overline{(B-V)}_\mathrm{Vmax} \approx$ 0.65 \& 1.04) when $^{56}$Ni is not
fully mixed ($f_\mathrm{m}$=0.15 \& 0.5), and the differences between them are
$\Delta \overline{(B-V)}_\mathrm{Vmax} \approx$ 0.20 and 0.16, which are also
comparable to the observed color difference.  Compared to the models with the
step $^{56}$Ni distributions, the models with Gaussian $^{56}$Ni distributions
were systematically redder for a given $f_\mathrm{m}$. The difference is most
prominent when $f_\mathrm{m}$=0.5, showing the color difference of $\Delta
\overline{(B-V)}_\mathrm{Vmax}\approx$ 0.45 between the step and the Gaussian
$^{56}$Ni distributions. This is because the $^{56}$Ni abundance in the outer layer of the ejecta is
higher in the models with a Gaussian distribution for a
given $f_\mathrm{m}$ value.  See Section~\ref{sec:mix} for a detailed
discussion on the effect of mixing.  On the other hand, the results with
$f_\mathrm{m} = 1.0$ are similar to the case of the step $^{56}$Ni
distributions because $^{56}$Ni is fully mixed throughout the ejecta for
both cases.

\section{Possible origins of the color difference between SNe Ib and Ic} \label{sec:col}

According to our models, there could be three possible reasons for the
systematic difference in the observed $(B-V)_\mathrm{Vmax}$ values of SNe Ib
and Ic.   
\begin{itemize}
\item Different $M_\mathrm{Ni}/M_\mathrm{ej}$ ratios between SNe Ib and Ic. 
\item Different amounts of helium  in the SNe Ib/Ic progenitors. 
\item Different degrees of  $^{56}$Ni mixing in the SNe Ib/Ic ejecta. 
\end{itemize}

Here we discuss each possibility in detail.

\begin{figure*}
\centering
\centering
\includegraphics[width=0.45\linewidth]{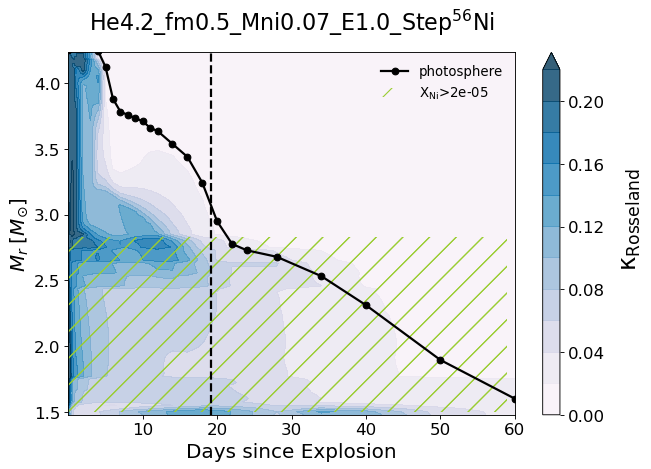}
\centering
\includegraphics[width=0.45\linewidth]{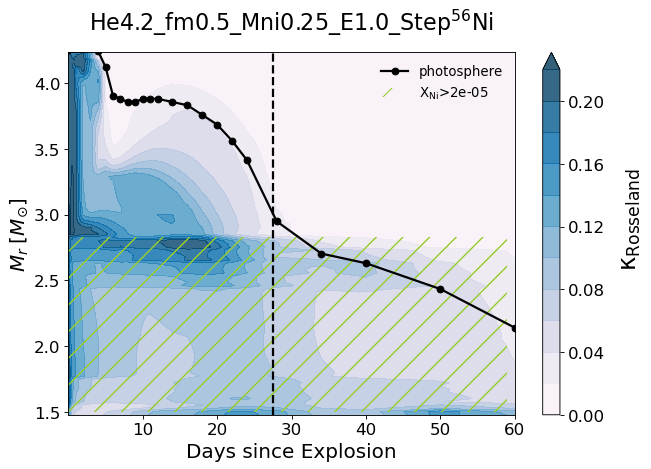}
\centering
\includegraphics[width=0.45\linewidth]{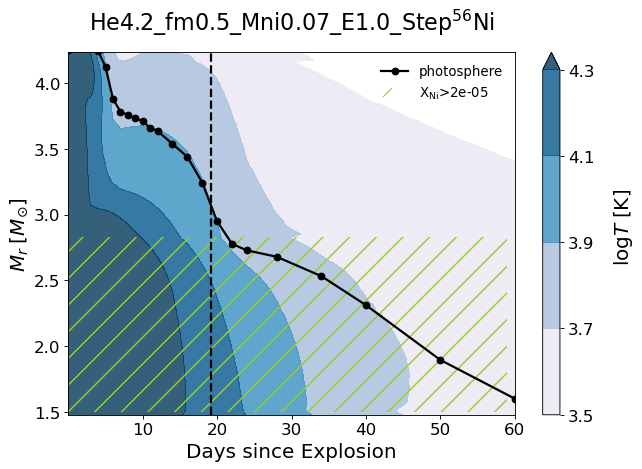}
\centering
\includegraphics[width=0.45\linewidth]{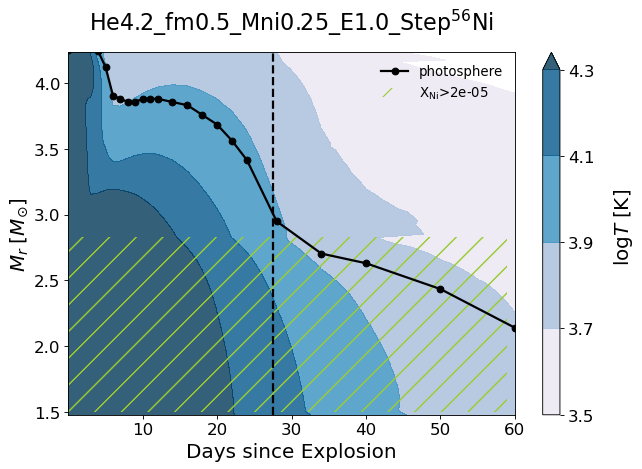}
\caption{The Rosseland-mean opacity evolution and the gas temperature evolution of the models for two different $^{56}$Ni mass: $M_\textrm{Ni}$=0.07$M_\odot$ (left panels) and 0.25$M_\odot$ (right panels). The models have a step $^{56}$Ni distribution. The vertical dashed line represents the $V-$band peak epoch.}
\label{fig:mni}
\end{figure*}

\begin{figure}
    \centering
    \includegraphics[width=0.45\textwidth]{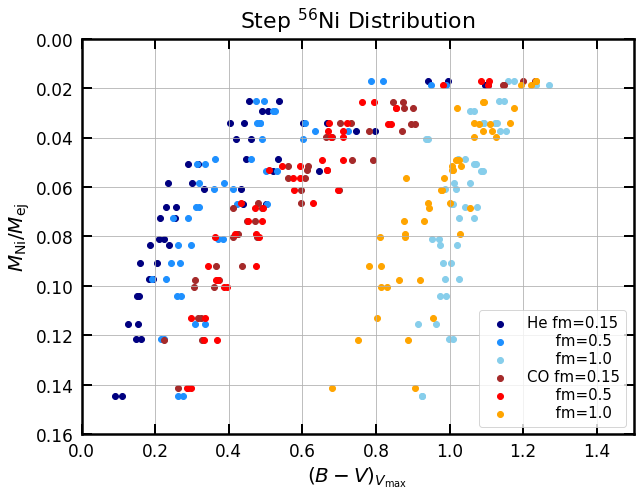}
    \includegraphics[width=0.45\textwidth]{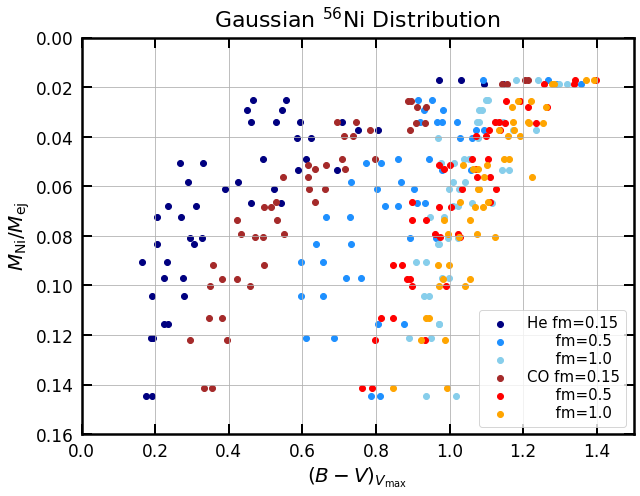}
    \includegraphics[width=0.45\textwidth]{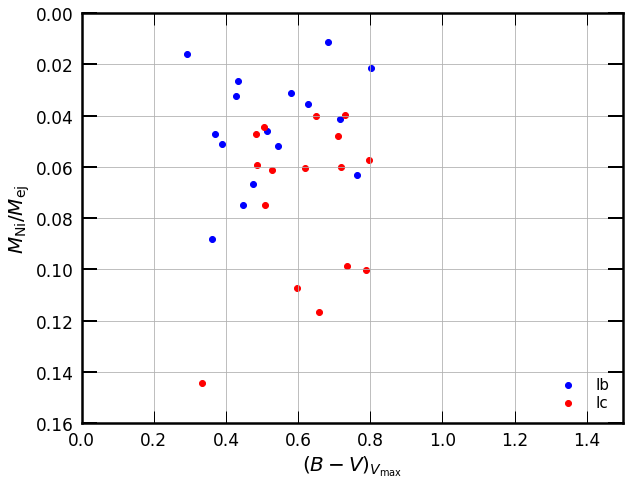}
\caption{The dependency of $(B-V)_\mathrm{Vmax}$ on $M_\textrm{Ni} / M_\textrm{ej}$. The SN models with the step (top panel) and 
Gaussian (middle panel) $^{56}$Ni distributions are the same models which are presented in Figure~\ref{fig:bvdist}. The data from our SN Ib/Ic sample are also presented (bottom). Mind that the y-axis is flipped.}
    \label{fig:MniMej}
\end{figure}

\begin{figure*}
\centering
\includegraphics[width=0.45\linewidth]{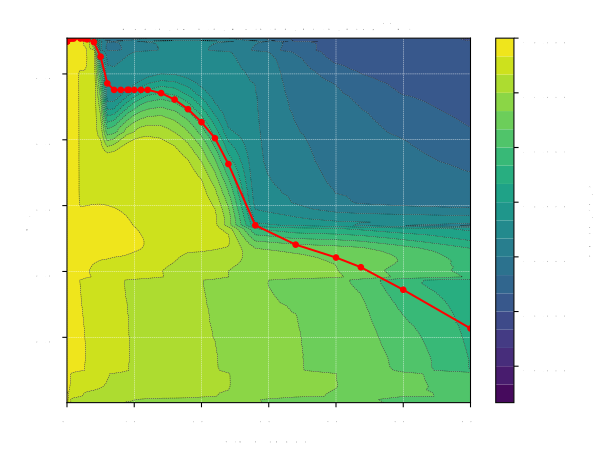}
\centering
\includegraphics[width=0.45\linewidth]{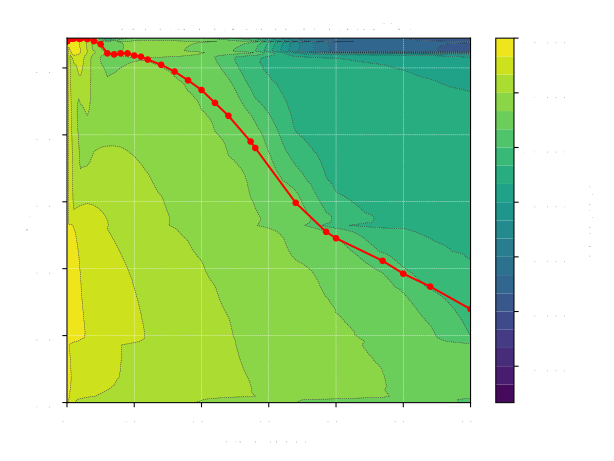}
\centering
\includegraphics[width=0.45\linewidth]{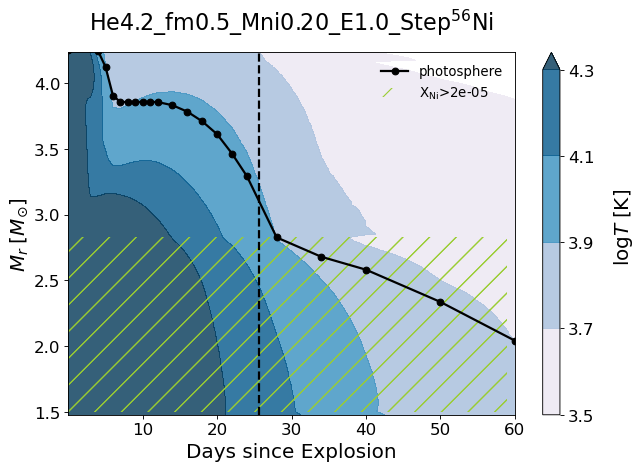}
\centering
\includegraphics[width=0.45\linewidth]{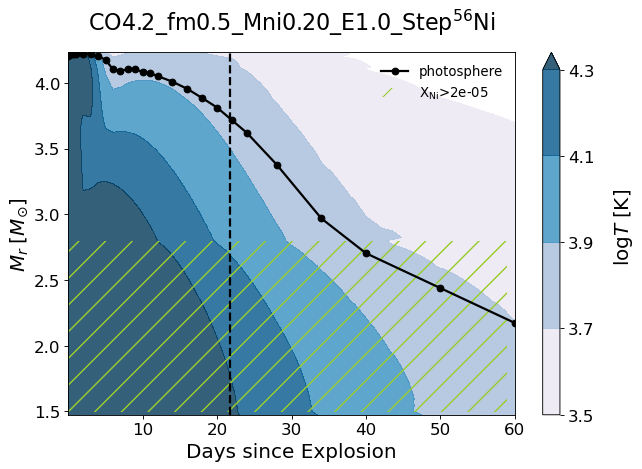}
\caption{The  evolution of the electron to baryon number density and the gas temperature evolution of the models with different chemical structures. The models have step $^{56}$Ni distribution. Left panels correspond to a He model, right panels correspond to a CO  model. Vertical dashed line represents the $V-$band peak epoch.
}
\label{fig:pro}
\end{figure*}

\begin{figure*}
\centering
\includegraphics[width=0.45\linewidth]{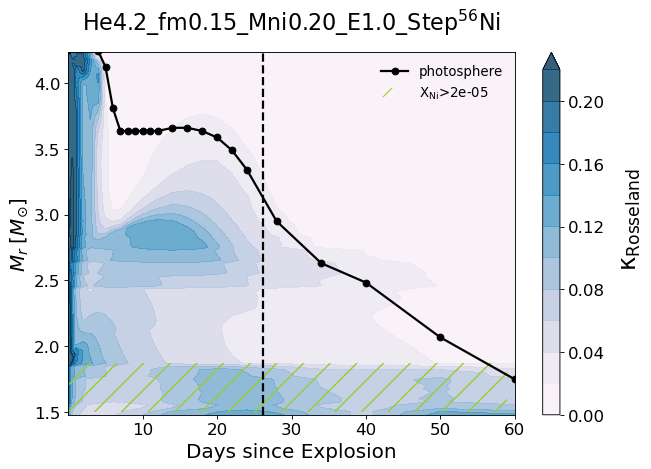}
\centering
\includegraphics[width=0.45\linewidth]{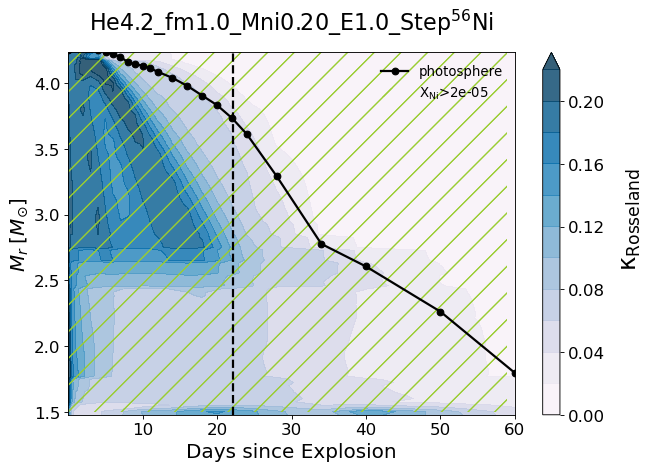}
\centering
\includegraphics[width=0.45\linewidth]{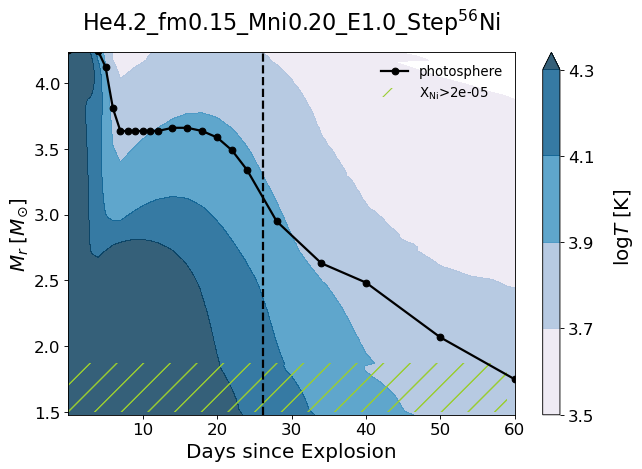}
\centering
\includegraphics[width=0.45\linewidth]{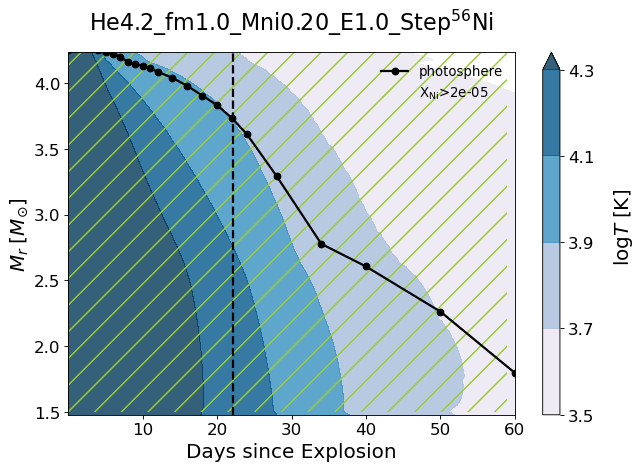}
\caption{The Rosseland-mean opacity evolution and the gas temperature evolution of the models with different degrees of $^{56}$Ni mixing. The models have step $^{56}$Ni distribution. Left panels correspond to $f_\mathrm{m}$=0.15, right panels correspond to $f_\mathrm{m}$=1.0. Vertical dashed line represents the $V-$band peak epoch.}
\label{fig:mix}
\end{figure*}

\begin{figure*}
\centering
\includegraphics[width=0.45\linewidth]{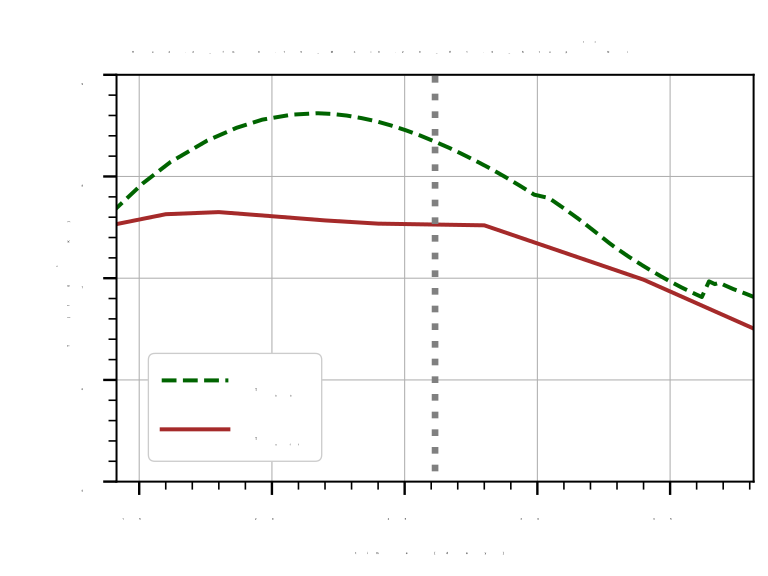}
\centering
\includegraphics[width=0.45\linewidth]{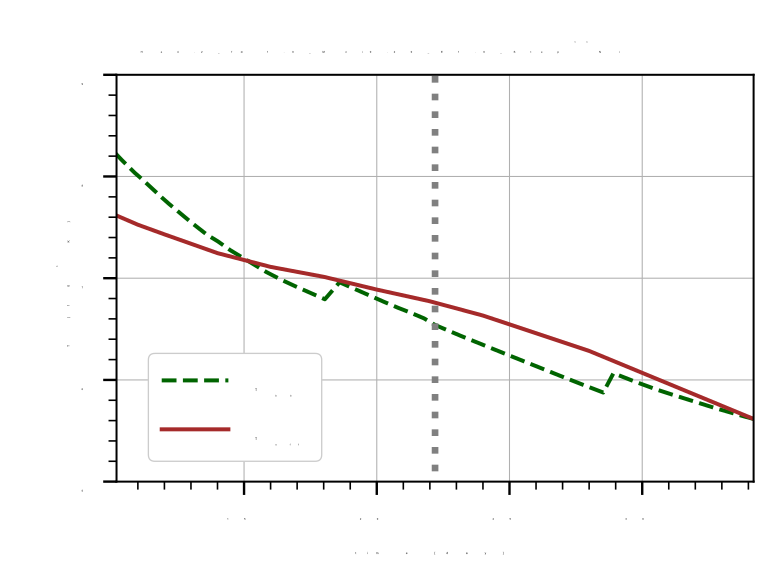}
\caption{Evolution of the effective temperature ($T_\mathrm{eff}$; solid line) at the photosphere
determined by the Rosseland-mean opacity and the temperature from the black-body fit to the SN spectrum ($T_\mathrm{bb}$;
dashed line)
in the He4.2 models 
with $M_\mathrm{Ni} = 0.20 M_\odot$ and $E_\mathrm{K} = 1.0$~B for two 
different step   $^{56}$Ni distributions of $f_\mathrm{m} = 0.15$ (left panel) 
and $f_\mathrm{m} = 1.0$ (right panel).
The vertical dotted line marks the epoch when the $V-$band peak is reached.
} 
\label{fig:Tbb}
\end{figure*}

\subsection{Ratio of $^{56}$Ni mass to ejecta mass}\label{sec:MniMej}


SN ejecta with more $^{56}$Ni for given ejecta mass and explosion energy would
have a larger thermal energy due to the extra heating.  Figure~\ref{fig:mni}
compares the evolution of the Rosseland-mean opacity and gas temperature in the
ejecta of the He4.2 model with $f_\mathrm{m} = 0.5$ for two different $^{56}$Ni
masses, $M_\mathrm{Ni}$=0.07$M_\odot$ and 0.25$M_\odot$. The ejecta with
$M_\mathrm{Ni}$=0.25$M_\odot$  is hotter at every mass zone for a given epoch
compared to the case of $M_\mathrm{Ni}$=0.07$M_\odot$. Accordingly, the ejecta
is more opaque because of more free electrons in the outer layers of the ejecta,
and the photosphere retreats more slowly. 
The photospheric gas temperature remains
in between $3.9<\log T[\mathrm{K}]<4.1$ at $t = 8
- 29$~d.  When the $V$-band peak appears at $t=27.5$~d,  the effective temperature
  at the Rosseland-mean photosphere ($T_\mathrm{eff}$) and the black-body fit temperature
of the SN spectrum ($T_\mathrm{bb}$),  which is more relevant to the SN color
than the effective temperature, 
are $T_\mathrm{eff} = 7820$~K and $T_\mathrm{bb} = 8320$~K,
respectively.  The corresponding $B-V$ is $(B-V)_\mathrm{Vmax} = 0.22$.  On the
other hand, The photospheric temperature of the $M_\mathrm{Ni} = 0.07 M_\odot$
model stays in $3.9<\log T[\mathrm{K}]<4.1$ at $t = 11 - 19$~d, duration of which
is shorter than the  $M_\mathrm{Ni} = 0.25 M_\odot$ case.  At the $V$-band peak
($t = 19.2$~d),  the effective and black-body fit temperatures are $T_\mathrm{eff} =
7370$~K and $T_\mathrm{bb} = 7240$~K, respectively.  The corresponding $B-V$ is
$(B-V)_\mathrm{Vmax} = 0.47$.  This comparison illustrates that a higher
$M_\mathrm{Ni}$ to $M_\mathrm{ej}$ ratio leads to a bluer optical color at
the $V$-band peak for a given initial condition. 


The dependency of $(B-V)_\mathrm{Vmax}$ on $M_\textrm{Ni} / M_\textrm{ej}$ can
also be observed in Figure~\ref{fig:MniMej}. As $M_\textrm{Ni} / M_\textrm{ej}$
becomes larger, $(B-V)_\mathrm{Vmax}$ of the models tends to decrease for a
given $^{56}$Ni distribution (note that the $y$-axis in the figure is flipped). 
By comparing the model color distributions
presented in Figures~\ref{fig:bvdist}
with the top and middle panels of Figure~\ref{fig:MniMej}, we can find that the
spread of $(B-V)_\mathrm{Vmax}$ for a given progenitor model and $^{56}$Ni
distribution mainly originates from the $M_\textrm{Ni} / M_\textrm{ej}$ spread.
For example, the $(B-V)_\mathrm{Vmax}$ distribution of the  CO models with a
step $^{56}$Ni distribution of $f_\mathrm{m}$=0.5 has the red end of the color distribution
($(B-V)_\mathrm{Vmax}$=1.1) when the models have the smallest $M_\textrm{Ni} /
M_\textrm{ej}$=0.02 and the blue end of the color distribution
($(B-V)_\mathrm{Vmax}$=0.3) when the models have the largest $M_\textrm{Ni} /
M_\textrm{ej}$=0.14.


It seems that the observed SN Ib/Ic sample does not follow the relation between
$M_\textrm{Ni}/ M_\textrm{ej}$ and $(B-V)_\mathrm{Vmax}$  predicted by the
STELLA models (see the bottom panel of Figure~\ref{fig:MniMej}). This might be
due to the limited sample size, the lack of $^{56}$Ni mixing information, poor
estimates of ejecta/explosion parameters, or host extinctions. Future extensive
investigation into SNe Ib/Ic photometry would test the validity of our model
prediction.


If $M_\mathrm{Ni} / M_\mathrm{ej}$ were systematically larger in SNe Ib than
SNe Ic, it could explain the systematic bluer color of SNe Ib than SNe
Ic of our SN sample (the middle panel of Figure~\ref{fig:obs}).  In contrast to
this expectation, however, it seems that SNe Ic have systematically larger
$M_\mathrm{Ni} / M_\mathrm{ej}$ than SNe Ib in our SN sample (the bottom panel
of Figure~\ref{fig:obs}).  
The same trend is also found 
in the most recent dataset of SNe Ib/Ic provided by \citet{Zheng2022}. 
We therefore conclude that the color
difference between SNe Ib and Ic cannot be attributed to different
$M_\mathrm{Ni}$ to $M_\mathrm{ej}$ ratios. 
However, given the large uncertainty in the estimates of $M_\mathrm{Ni}$ and $M_\mathrm{ej}$
as discussed in Section~\ref{sec:obs}, this conclusion should be only considered tentative.

\subsection{Helium contents in the progenitor}\label{sec:hemass}

According to the most popular SN scenario, SNe Ib and Ic progenitors are
helium-rich and helium-poor, respectively~\citep[e.g.,][]{Yoon2015}.  This scenario has been supported by
many observational and theoretical studies as discussed in
Section~\ref{sect:introduction}~\citep[e.g.,][]{Matheson2001, Dessart2012,
Hachinger2012, Liu2016, Fremling2018, Yoon2019, Dessart2020, Moriya2020,
Teffs2020, Williamson2020, Shahbandeh2022}.  Recent stellar evolution models
also suggest that SN Ib and Ic progenitors would not form a continuous sequence
in terms of the He amount if mass-loss enhancement during the WC phase of WR
stars, which is implied by recent WR star observations of the local universe,
is assumed~(see \citealt{Yoon2017b} for detailed discussion and references
therein; see also \citealt{Woosley2021} and \citealt{Aguilera2022}).  In this
Section, we explore the consequence of the dichotomy nature of helium contents
in the SN Ib/Ic color~\citep[see also][]{Yoon2019}.   

In Figure~\ref{fig:pro}, we present the evolution of the ratio of the free
electron to baryon numbers ($n_\mathrm{e}/n_\mathrm{b}$) and the gas
temperature  in the ejecta of the He4.2 and CO4.2 models with $E_\mathrm{K} =
1.0$~B, $M_\mathrm{Ni} = 0.2~M_\odot$ and a step $^{56}$Ni distribution
($f_\mathrm{m} = 0.5$).  In the He4.2 model, the ratio
$n_\mathrm{e}/n_\mathrm{b}$ in the outermost layers (i.e., $M_r \gtrsim 4.0
M_\odot$) rapidly decreases after $t \simeq 4$~d as the temperature decreases and
the Rosseland-mean photosphere rapidly moves inwards accordingly in the mass-coordinate.
However, in the CO4.2 model, more free electrons are available in the outer
layers of the ejecta and  the decrease of $n_\mathrm{e}/n_\mathrm{b}$ is
relatively slow compared to the He4.2 model.  This makes the photosphere of the
CO4.2 model retreat inwards more slowly than in the He4.2 model: $M_{r,\mathrm{ph}}
[M_\odot] = 4.2 \rightarrow 3.9 \rightarrow 2.8$ (He4.2) and $M_{r,
\mathrm{ph}} [M_\odot] = 4.2 \rightarrow 4.1 \rightarrow 3.4$ (CO4.2) at $t=0$~d
$\rightarrow$ 6 d $\rightarrow$ 28 d, respectively.  The photospheric gas
temperature at $t = 5
- 28$ d is higher in He4.2 model ($3.9 < \log T[\mathrm{K}] < 4.1$) than in
  CO4.2 model ($ 3.7 < \log T[\mathrm{K}] < 3.9$). 

The $V-$band peaks are reached when $t = 25.6$~d and $t = 21.8$~d for the He4.2 and
CO4.2 models, and the corresponding effective and black-body fit temperatures are
$T_\mathrm{eff}=7610$~K and $T_\mathrm{bb}=8234$~K for He4.2 and
$T_\mathrm{eff}=6774$~K and $T_\mathrm{bb}=7327$~K for CO4.2, respectively. 

For this reason, He models are systematically bluer than CO  models for a given
set of model parameters except for the (almost) full $^{56}$Ni-mixing cases
(i.e., $f_\mathrm{m} = 1.0$; Figure~\ref{fig:bvdist}).  The
effect of $^{56}$Ni  is discussed in the section that follows.  As shown in
Figure~\ref{fig:bvdist}, the differences in
$(B-V)_\mathrm{Vmax}$  between the two progenitor types (i.e., He and CO) in
our models are $\Delta (B-V)_\mathrm{Vmax} \approx$ 0.14-0.18 for a given
$^{56}$Ni distribution, which is comparable to the $\Delta
(B-V)_\mathrm{Vmax}$ value for our observation sample of SNe Ib/Ic.
This is in good agreement with the notion that SNe Ib originate from
helium-rich progenitors and SNe Ic from helium-poor progenitors.

\subsection{$^{56}$Ni mixing}\label{sec:mix}


Radioactive $^{56}$Ni can significantly affect the SN color by radioactive
heating, ionization induced by the heating, and line blanketing.
Although a higher $M_\mathrm{Ni}/M_\mathrm{ej}$ leads to a bluer color at the
optical peak for a given $^{56}$Ni distribution as discussed in
Section~\ref{sec:MniMej}, we find that a stronger $^{56}$Ni mixing can make the
color redder at the $V$-band peak for a given $M_\mathrm{Ni}/M_\mathrm{ej}$
ratio, as shown in Figure~\ref{fig:bvdist}: the fully mixed case
($f_\mathrm{m}$=1.0) results in a much redder color
($\overline{(B-V)}_\mathrm{Vmax}$ $\approx 1.0$) than weakly/moderately mixed
cases ($f_\mathrm{m}$=0.15, 0.5, $\overline{(B-V)}_\mathrm{Vmax}$ $\approx 0.4
- 0.6$). 

In Figure~\ref{fig:mix}, we show the Rosseland-mean opacity and temperature
evolution in the ejecta of the He4.2 models for $f_\mathrm{m} = 0.15$ (weak $^{56}$Ni mixing)
and 1.0 (full $^{56}$Ni mixing) with the step $^{56}$Ni distributions.  It is
observed that in the case of $f_\mathrm{m} = 0.15$,  the $^{56}$Ni heating  of
the outer layers is somewhat delayed, while the ejecta with $f_\mathrm{m} = 1.0$ monotonically cools down 
given that $^{56}$Ni is evenly distributed throughout the ejecta.  Compared to
the $f_\mathrm{m}$=1.0 case, the Rosseland-mean photosphere in the
$f_\mathrm{m}$=0.15 model retreats more rapidly but stays in the hot
$^{56}$Ni-heated region ($\log T[\mathrm{K}] > 3.9$ at $t=11 - 30$~d) during which
the $V$-band peak appears at $t = 26$~d. 
The Rosseland-mean photosphere in the
$f_\mathrm{m}$=1.0 model recedes inwards more slowly and reaches the $V$-band
peak earlier (i.e., at $t = 22.1$~d). This is because of more free electrons in
the outer layers due to the  $^{56}$Ni heating and line blanketing due to the
presence of $^{56}$Ni.  This leads to lower effective and black-body fit
temperatures for a stronger $^{56}$Ni mixing at the $V$ band peak:
$T_\mathrm{eff} = 7528$~K and $T_\mathrm{bb} = 83420$~K for $f_\mathrm{m} =
0.15$ and   $T_\mathrm{eff} = 6760$~K and $T_\mathrm{bb} = 6540$~K for $f_\mathrm{m} = 1.0$.

The effect of line blanketing can also be observed by comparing the evolution
of $T_\mathrm{eff}$ and $T_\mathrm{bb}$ as in Figure~\ref{fig:Tbb}.  In the
case of $f_\mathrm{m} = 0.15$, the Rosseland-mean photosphere remains in
$^{56}$Ni-free layers and $T_\mathrm{bb}$ is systematically higher than the
effective temperature at the Rosseland-mean photosphere during the epochs
around the $V-$band peak. This is because electron scattering causes the
thermalisation depth to be located below the Rosseland-mean photosphere. The
monochromatically scattered photons keep $T_\mathrm{bb}$ corresponding to those
deep layers. Their  number density is reduced by the dilution effect and
the result is a bluer spectrum than the black-body spectrum corresponding to
$T_\mathrm{eff}$. By contrast, $T_\mathrm{bb}$ is lower than $T_\mathrm{eff}$
in the $f_\mathrm{m} = 1.0$ case because line blanketing due to $^{56}$Ni
obscures photons having short wavelengths and this is mainly responsible for
the redder SN color with a stronger $^{56}$Ni mixing.

We find that the colors of the Gaussian $^{56}$Ni distribution models with weak
and moderate $^{56}$Ni mixing ($f_\mathrm{m} =$ 0.15 and 0.5) are
systematically redder than the corresponding step $^{56}$Ni distribution
models~(see Figure~\ref{fig:bvdist} and
Table~\ref{tab:avg}).  This result can also be explained by the effects of
$^{56}$Ni on the SN spectral energy distribution.  As seen in
Figure~\ref{fig:abn}, the $^{56}$Ni distribution following the Gaussian
function extends to the outermost layers even for the case of $f_\mathrm{m} =$
0.15 and 0.5, while with the step function,  $^{56}$Ni is present only in the
innermost confined region unless $^{56}$Ni is fully mixed.  Therefore, in the
models with the Gaussian $^{56}$Ni distribution,   $^{56}$Ni is always present
at the Rosseland-mean photosphere and the resulting line blanketing makes the
SN color redder than the corresponding step distribution models.

We conclude that different degrees of $^{56}$Ni mixing would make the
$(B-V)_\mathrm{Vmax}$ color different even when their progenitor types are the
same in terms of helium content. However, it is not likely 
that the color difference between SNe Ib and Ic in our observation sample 
is soley due to this mixing effect. 

\citet{Yoon2019} argue that the non-monotonic and monotonic color evolution
observed in SNe Ib and some SNe Ic during early times implies relatively weak
and very strong  $^{56}$Ni mixing in SN Ib and Ic ejecta, respectively.
However, a stronger $^{56}$Ni mixing into the helium-rich layer implies
formation of strong \ion{He}{1} lines during the photospheric
phase~\citep{Dessart2012, Hachinger2012, Dessart2020, Teffs2020,
Williamson2020}.  This means that if the redder color of SNe Ic compared to SNe
Ib were mainly due to a stronger $^{56}$Ni mixing, the SN Ic progenitors must
be helium-poor.  As shown in Section~\ref{sec:hemass}, on the other hand,
helium deficiency would also lead to a redder color compared to the helium-rich
case for a given $^{56}$Ni distribution.  Therefore, it is possible that, in
reality, the redder color of SNe Ic compared to SNe Ib is related to both
helium-deficiency and a stronger $^{56}$Ni mixing.

\section{Conclusions} \label{sec:con}

We show that the optical colors of observed SNe Ib and SNe Ic are systematically different
at the $V$-band peak in our selected sample (16 SNe Ib and 17 SNe Ic;
Table~\ref{tab:obs}): SNe Ib are bluer ($\overline{(B-V)}_\mathrm{Vmax}=0.52$)
than SNe Ic ($\overline{(B-V)}_\mathrm{Vmax}=0.62$) by $\Delta
\overline{(B-V)}_\mathrm{Vmax}=$ $0.10$, on average (Figure~\ref{fig:obs}; Section~\ref{sec:obs}). 
This color difference is found to be larger (i.e., $\Delta\overline{(B-V)}_\mathrm{Vmax}= 0.22$)
if we limit our sample to the minimally reddened case (i.e., $E(B-V)_\mathrm{host} \le 0.05$).  

Using multi-band radiation-hydrodynamics simulations with the STELLA code for
both helium-rich and helium-poor progenitors of various final masses ($M \simeq
3.1 \cdots 5.7 M_\odot$), we explore three possible reasons for the color
difference: 1) different $M_\mathrm{Ni}/M_\mathrm{ej}$ ratios
(Section~\ref{sec:MniMej}), 2) different amounts of helium
(Section~\ref{sec:hemass}), and 3) different degrees of $^{56}$Ni mixing in SN
ejecta (Section~\ref{sec:mix}).  We find that the SN color becomes bluer at the
$V-$band peak for a larger $M_\mathrm{Ni} / M_\mathrm{ej}$ ratio, a weaker
$^{56}$Ni mixing, and/or a helium-rich progenitor compared to the corresponding
helium-poor case~(Figure~\ref{fig:bvdist}~ and
\ref{fig:MniMej}). 

From these results, we draw the following conclusions:
\begin{enumerate}
\item In our sample of observed SNe, the
$M_\mathrm{Ni}/M_\mathrm{ej}$ ratios in SNe Ic seem to be systematically higher
than in SNe Ib (see Figure~\ref{fig:obs}). This implies that if the inner
structure and the degree of $^{56}$Ni mixing in SNe Ib and SNe Ic were similar to
each other, SNe Ic would be systematically bluer than SNe Ib, in sharp contrast
to the observation. Therefore, we conclude that different $M_\mathrm{Ni}/M_\mathrm{ej}$ ratios 
cannot explain the color difference between SNe Ib and Ic (Section~\ref{sec:MniMej}). 
\item We also exclude the possibility that radioactive $^{56}$Ni is almost fully
mixed in both SNe Ib and SNe Ic ejecta, as otherwise no systematic $B-V$ color
difference would be observed~(see Figure~\ref{fig:bvdist}). 
\item We find that the $B-V$ color difference can be well explained by the standard scenario for SNe Ib and Ic (i.e., helium-rich 
and helium-poor progenitors for SNe Ib and SNe Ic, respectively), given that 
the color at the $V-$band peak is systematically redder with the helium-poor progenitors for a given  $^{56}$Ni distribution (unless $^{56}$Ni is fully mixed).  
It is possible that the redder color of SNe Ic are partly due to a stronger $^{56}$Ni mixing in the ejecta, compared to SNe Ib~(Section~\ref{sec:mix}). 
If this is the case, SNe Ic progenitors must be helium-poor as otherwise strong \ion{He}{1} absorption lines would be
detected in SN Ic spectra.  
\end{enumerate}

In short, we conclude that the systematic color difference between SNe Ib and
SNe Ic at the $V-$band peak provides strong evidence for the distinct
properties of their progenitors in terms of helium content, rebutting the
existence of a large amount of hidden helium in SNe Ic. 

This study is subject to a few limitations. First of all, STELLA might not
predict broad-band colors at the optical peak accurately due to some physical
simplifications implemented in the code, e.g., LTE approximation for atomic
level populations, the limited number of spectral lines, the lack of proper
treatment of the fluorescent effect, etc. Detailed spectrum calculations
including these factors would be required for more rigorous and quantitative
comparison of the models with the observation.  Secondly, only a small number
of SNe Ib/Ic (16/17 for each SN subtype) are used for the analysis because not
many sufficiently good photometric or host extinction data are available.
Future acquisition of large samples of SN Ib/Ic will allow us to confirm the
results obtained in the study.  

Finally, our model comparison with the
observations is limited to the $B-V$ color at the optical peak. It would be
interesting to extend our approach to post-maximum
colors, 
which might provide further constraints on the nature of SNe Ib/Ic progenitors
and the SN ejecta properties
~\citep[e.g.,][]{Drout2011, Dessart2015, Dessart2016, Stritzinger2018, Woosley2021}. 
However, the SN Ib/Ic colors during the
post-maximum can be much more significantly affected by specific lines and non-LTE
effects than earlier times~\citep[e.g.,][]{Dessart2015, Dessart2016}, which are not properly
considered in the current version of STELLA. The time-dependent non-LTE
effects are being implemented in STELLA and a more extended study on the SN
Ib/Ic colors including post-maximum colors will be presented in the future.

\acknowledgments

Wa thank the referee for helping us improve the manuscript. This work has been
supported by the National Research Foundation of Korea (NRF) grant
(NRF-2019R1A2C2010885).  We are grateful to Taebum Kim for creating the python
package for the Kippenhahn diagram and to Wonseok Chun for providing the
progenitor models to us. Work by S.B. on the development of STELLA code is
supported by the Russian Science Foundation grant 19-12-00229 and by RFBR
21-52-12032 on SNIc studies. 

\bibliography{bib}{}

\appendix
\section{appendix section}

\begin{table*}[]
\caption{$M_\mathrm{Ni}$ and $M_\mathrm{ej}$ of our selected SN Ib/Ic sample}\label{tab:MniMej}
\begin{center}
\begin{tabular}{ccc}
\hline \hline
Name & $M_\mathrm{Ni}$ [$M_\odot$] & $M_\mathrm{ej}$ [$M_\odot$] \\ \hline
SN 1999ex & 0.25 (R06), 0.1 (D11), 0.12 (C13), 0.15 (L16), 0.172 (P16) & 0.9 (R06), 2.91 (C13), 2.9 (L16) \\
SN 2004gq & 0.13 (D11), 0.14 (C13), 0.1 (L16), 0.11 (T18) & 3.19 (C13), 1.8 (L16), 3.4 (T18) \\
SN 2004gv & 0.14 (C13), 0.16 (T18) & 11.72 (C13), 3.4 (T18) \\
SN 2006ep & 0.06 (L16), 0.12 (T18) & 2.7 (L16), 1.9 (T18) \\
SN 2006gi & 0.064 (E11) & 3.0 (E11) \\
SN 2006lc & 0.3 (T15), 0.14 (T18) & 3.67 (T15), 3.4 (T18) \\
SN 2007C & 0.16 (D11), 0.18 (C13), 0.17 (L16), 0.07 (T18) & 1.83 (C13), 1.9 (L16), 6.2 (T18) \\
SN 2007kj & 0.066 (T18) & 2.5 (T18) \\
SN 2007Y & 0.03 (C13), 0.04 (L16), 0.051 (P16), 0.03 (T18) & 2.09 (C13), 1.4 (L16), 1.9 (T18) \\
SN 2008D & 0.07 (D11), 0.08 (C13), 0.09 (L16), 0.111 (P16) & 5.33 (C13), 2.9 (L16) \\
SN 2009jf & 0.18 (C13), 0.24 (L16), 0.271 (P16) & 7.34 (C13), 4.7 (L16) \\
SN 2012au & 0.3 (M13) & 4.0 (M13) \\
SN 2014C & 0.15 (M17) & 1.7 (M17) \\
SN 2015ah & 0.092 (P19) & 2.0 (P19) \\
SN 2015ap & 0.12 (P19) & 1.8 (P19) \\
iPTF13bvn & 0.06 (L16), 0.07 (P16) & 1.7 (L16) \\ \hline
SN 1994I & 0.08 (R06), 0.06 (D11), 0.06 (C13), 0.07 (L16), 0.102 (P16) & 0.5 (R06), 0.72 (C13), 0.6 (L16) \\
SN 2004aw & 0.27 (D11), 0.22 (C13), 0.2 (L16) & 6.49 (C13), 3.3 (L16) \\
SN 2004dn & 0.16 (D11), 0.15 (C13), 0.16 (L16) & 3.4 (C13), 2.8 (L16) \\
SN 2004fe & 0.19 (D11), 0.19 (C13), 0.23 (L16), 0.1 (T18) & 2.07 (C13), 1.8 (L16), 2.5 (T18) \\
SN 2004gt & 0.16 (T18) & 3.4 (T18) \\
SN 2005aw & 0.17 (T18) & 4.3 (T18) \\
SN 2007gr & 0.07 (D11), 0.04 (C13), 0.08 (L16), 0.073 (P16) & 1.7 (C13), 1.8 (L16) \\
SN 2007hn & 0.25 (T18) & 1.5 (T18) \\
SN 2011bm & 0.58 (C13), 0.62 (L16), 0.702 (P16) & 18.75 (C13), 10.1 (L16) \\
SN 2013F & 0.15 (P19) & 1.4 (P19) \\
SN 2013ge & 0.109 (P16) & 2.5 (D16) \\
SN 2014L & 0.075 (Z18) & 1.0 (Z18) \\
SN 2016iae & 0.13 (P19) & 2.2 (P19) \\
SN 2016P & 0.09 (P19) & 1.5 (P19) \\
SN 2017ein & 0.13 (X19) & 0.9 (X19) \\
SN 2020oi & 0.07 (R20) & 0.71 (R20) \\
LSQ14efd & 0.25 (J21) & 2.49 (J21) \\ \hline
\end{tabular}
\end{center}
\tablecomments{References are abbreviated as follows. R06: \citet{Richardson2006}, D11: \citet{Drout2011}, E11: \citet{Elmhamdi2011}, O12: \citet{Oates2012}, C13: \citet{Cano2013}, M13: \citet{Milisavljevic2013} median value adopted, M15: \citet{Milisavljevic2015}, T15: \citet{Taddia2015}, D16: \citet{Drout2016} median value adopted, L16: \citet{Lyman2016}, P16: \citet{Prentice2016} host corrected values, B17: \citet{Barbarino2017}, M17: \citet{Margutti2017}, S18: \citet{Stritzinger2018},  T18: \citet{Taddia2018} hydrodynamical model, V18: \citet{VanDyk2018}, Z18: \citet{Zhang2018}, P19: \citet{Prentice2019}, X19: \citet{Xiang2019}, R21: \citet{Rho2021}, J21: \citet{Jin2021}}
\end{table*}

\end{document}